# Fast Load Balancing Approach for Growing Clusters by Bioinformatics

Soumen Kanrar
Department of Computer Science
Vidyasagar University, Midnapour, WB India
rscs_soumen@mail.vidyasagar.ac.in

*Abstract*—This paper presents Fast load balancing technique inspired by Bioinformatics is a special case to assign a particular patient with a specialist physician cluster at real time. The work is considered soft presentation of the Gaussian mixture model based on the extracted features supplied by patients. Based on the likelihood ratio test, the patient is assigned to a specialist physician cluster. The presented algorithms efficiently handle any size and any numbers of incoming patient requests and rapidly placed them to the specialist physician cluster. Hence it smoothly balances the traffic load of patients even at a hazard situation in the case of natural calamities. The simulation results are presented with variable size of specialist physician clusters that well address the issue for randomly growing patient size.

## I. INTRODUCTION

At the current discard, people need better medical advice at real time. The major issue is to assign the patient to the specified specialist physician without any delay. In the previous work by Nan Liu el. al.[1] have considered the heuristic policies for scheduling patient appointments taking into account the fact that patients may cancel or defer their appointments. Nan Liu el al.[1] have considered various numbers of heuristic policies to present the scheduling model. A. Hertz and N. Lahrichi [2] have addressed the problem in a different way for assigning patients to nurses in the course of home-care services. A. Hertz and N. Lahrichi [2] addressed the workload balance of the nurses, to avoid long travel time for the visit of patients. In this regards, A. Hertz and N. Lahrichi [2] have proposed 'Tabu' search algorithm for the patient assignment problem. The 'Tabu' search algorithm given a solution space $X$ and a function $f$ that measures the value $f(x)$ of every solution $x \in S$, and $X \subset S$. Their proposed 'Tabu' search algorithm had a specific objective to determine a specific solution $x^*$, which is used to minimize $f(x^*)$ over $X$. The obtained minimum value is nothing, but the minimum load assigns to the nurse. The survey paper by Gupta and Denton [3, 4] vigorously focused on the practical issues related to appointment scheduling that provides a review of the state of modeling and optimization. Gupta and Denton [3, 4] addressed to the future directions regarding the necessity of bioinformatics in the area of load balancing. The classification made by Gupta and Denton [3, 4] regarding the research on appointment of scheduling with respect to the type of waiting modeled as direct and indirect. Gupta and Denton[3,4] indicate, most of the existing research has concentrated on direct waiting times. The direct waiting time is the time the patients generally considered to spend waiting in the clinic on that day of appointments. That work typically analyzed to minimize the expected "cost of time" for a day, which is a function of patient's direct waiting times, and the physician's idle time or overtime. Scott Levin et. al [5] have founded an important and apparent imbalance in the distribution of load balancing among all physicians working concurrently. Levin and France [6, 7] have considered the work load and communication patterns for individual physicians in emergency working during the periods of high demand. Still the issue is remained challenging one. A new type of approach is required to address the problem in an efficient way. Currently for the speaker identification and verification is done based on bioinformatics. In speaker identification and verification major two types of approach are considered one is Gaussian Mixture model [8, 9, 10]. Another approach is on the base of 'i-vector', that is nothing but space and dimension compactness of GMM generated space [11, 12]. Major issue is to handle a certain growth of the patient set. People seek 'various medical advices' from the specialist physician. Article [15] addresses the performance of clustering particularly in the mobile domain without considering the patients biological data and information. Particularly in the wireless medium i.e. the patient used to send their biological data and information, softly consider the handoff issue. The parametric estimation for handoff [13] will be considered as the case of those patients. Those are using the smart phone to send their biological data and information. Some cloud base approach can enhance the problem, particularly for private cloud job allocation [14]. The current issue is considered as how the patient assigns to the ideal specialist physician or fewer loaded specialist physicians. The Biological data and information [16] have a great impact for the acutely serious condition patient. The main goal of this work to be proper balancing the ever increase patient load uniformly according to their initial extracted basic feature parameters. The same concept can be further extended to distribute the traffic load into the different server according to the initial basic characteristic of the packet data type in a distributed system based on this bioinformatics concept. This paper is structured as follows. Section one introduces the problem. The basic survey related to load balancing of patient based on bioinformatics is presented in this section. Section two present the clustering formation of the specialist physicians as well as general physician. Section three present the model formation is based on the bioinformatics information supplied by the patient. Section four presents the two general algorithms related to patient allotment to physician cluster and recursively update of the physician accepted list. The result analysis of the simulation is presented in the section five with a conclusion at the end.

## II. CLUSTERING OF PHYSICIANS

The registered lists of physicians (in a society) are classified into the $k$ number of clusters. Here, $(k-1)$ numbers of clusters are specified for the specialized physician in $(k-1)$ specialized domain. The $k^{th}$ cluster is fixed for the general physician. The description of clusters is as follows. For example, the cluster skin contains the physician's expert in the skin domain. Let us consider that cluster as $C_1$. The cluster orthopedic contains the physicians related to the domain of orthopedic. Let us consider the cluster as $C_2$. The cluster ENT contains the physicians related to ENT. Let us consider the cluster as $C_3$. In this procedure, we are supposed to generate $(k-1)$ number of specialized physician clusters in the $(k-1)$ specialized field. The vital medical information related to each patient is collected by extracting the feature's vector from the submitted information by patient or the representative of the patient. The patient submitted all the biological and individual health information through the wired or wireless medium. Wired medium may be affected during the natural disaster but wireless medium is very effective during the natural disaster.

## III. MODEL FORMATION

Let us consider $X$ is a random vector. It is expressed as $X = \{x_1, x_2, \cdots, x_n\}$. So, $X$ be a set of n vectors each $x_i$ of is a $k$ dimensional *feature- vector* extracted from the submitted patient information. Those individual $x_i$ vectors are statistically independent. The probability distribution of the set X is based on the given model $\lambda$ expressed as

$$\log p(X/\lambda) = \sum_{h=1}^{n} \log p(x_h/\lambda) \qquad (1)$$

As usual the distributions of these vectors are unknown. So the soft presentation can be better approximated by a general model. The general model with respect to the variable weight for extracted feature is the mixture of Gaussian probability distributions. It is a weighted sum of $l$ component densities is expressed by the equation

$$p(x_h/\lambda) = \sum_{i=1}^{l} w_i N(x_h, \mu_i, \Sigma_i) \qquad (2)$$

$\lambda$ is the prototype consisting of a set of model parameters and express as $\lambda = \{w_i, \mu_i, \Sigma_i\}$, here $w_i$ is the mixture weight, with $\sum_{i}^{n} w_i = 1$ and $N(x_i, \mu_i, \Sigma_i)$ is the $n$ variate Gaussian components densities with mean vectors $\mu_i$ and covariance matrices $\Sigma_i$. The probability distribution for the extracted feature vectors i.e. for the random vector is

$$p(X/\lambda) = e^{\sum_{h=1}^{n} \sum_{i=1}^{l} w_i \frac{\exp\{-\frac{1}{2}(x_h - \mu_i)' \Sigma_i^{-1} (x_h - \mu_i)\}}{(2\Pi)^{k/2} |\Sigma_i|^{\frac{1}{2}}}}$$

Here, $0 \leq x_h \leq \infty$ and $1 \leq i \leq n$, $(x_h - \mu_i)'$ is the transpose of $(x_h - \mu_i)$ and $\Sigma_i^{-1}$ is the inverse of the covariance matrix $\Sigma_i$. The Standard GMM model $\{G_1, G_2, \cdots, G_{k-1}\}$ related to the Physician cluster $\{C_1, C_2, \cdots, C_{k-1}\}$ i.e. $G_i$ model represents the cluster $C_i$. Let, $G^*$ is the GMM (Gaussian Mixture Model) for individual patients submitted information through the wired or wireless medium. Here we find the association of the patient to a particular specialist physician cluster by the likelihood ratio test based on a hypothesis. The likelihood function is used to reserving the roles of the data vector and parameter vector. Here we consider $L(\lambda/X) \approx P(X/\lambda)$ for the required algorithm development part. Two major types of algorithm are developed based on the extracted feature vector. The algorithm assigns patient to the specified specialist physician cluster and the second algorithm recursively updates the physician list during the patient assignment to the particular physician cluster.

## IV. ALGORITM DEVELOPMENT

The algorithm one presents the allotment of the incoming patient to the specialist physician cluster (according to the submitted information via the wired or wireless mediums).

| Algorithm 1: Patient allotment |
|---|
| 1. **Var** |
| 2. $i$ : Integer |
| 3. Trial: Boolean |
| 4. Threshold: Real |
| 5. $p$ : Character string // unique Patent ID assign by //system |
| 6. **Begin** |
| 7. $i = 1$ |

8.     While (Trial= =true) do
9.     If $\frac{L(\lambda_*/X)}{L(\lambda/X)} \approx \frac{\log p(X/\lambda_*)}{\log p(X/\lambda_i)} \geq$ threshold

// $\lambda_*$ is the model parameter for incoming patient information

// $\lambda_i$ is the standard model parameter for the $i^{th}$ cluster

10.     **Allot:** Patient $p$ assign to Cluster $C_i$
11.     then  Trial = False
12.     **End if**
13.     If ( $i = = k$ )
14.     **Allot:** Patient $p$ assign to Cluster $C_k$  // General                                                          // physician
15.     Trial = False
17.     End if
18.     Else
19.     $++i$
20.     **End while**
21.     End

The algorithm two present the recursive procedure call that update the specialist physician accept list of patient.

Algorithm 2: Recursive Update of Accepted list

1.     **Var**
2.     T, t, $\Delta t$, $X_y$ : real
3.     Number_ Patent_Assign _To_ Physician: integer
4.     At T = t // System time

// Assign patient P to Physician $d_i^j \in C_i$ i.e. Least loaded,

// $j^{th}$ physician in $i^{th}$ cluster

5.     **Module Update**:

// Find the minimum load of the physician for the $i^{th}$ cluster

6.     $v \leftarrow$ Size of Cluster ( $C_i$ )
7.     $j \leftarrow 1$
8.     While $(j \neq v)$ do
9.     if ( $d_i^j$ .Number_ Patent_Assign _To_Physician
           $< d_i^{j+1}$. Number_ Patent_Assign _To_Physician)
10.    Min_Load_Physician $\leftarrow d_i^j$

// Find Minimum loaded physician,

// $d_i^j$ is a logical identifier number

11.    **Else**
12.    Min_Load_Physician $\leftarrow d_i^{j+1}$
13.    $j \leftarrow j+1$
14.    **End while**

// Patient admit list add P to Physician, in a particular cluster
//i.e. the physician of a particular cluster with minimum load
//be updated

15.    Min_Load_Physician $\leftarrow$ Min_Load_Physician $\cup \{P\}$

// Find Mean of the physician load at time T=t

// i.e. with picked up duration is $\Delta t$ (system defined)

16.    At T = t + $\Delta t$
17. $E(X_y) \leftarrow \frac{1}{sizeof(C_y)} \sum_{j=1}^{s} (d_{i=y}^j .\text{Number\_Patent\_Assign}$
                              _To_Physician)
18.    If $\| E(X_y)_{t=t} - E(X_y)_{t=t+\Delta t} \| \approx 0$ then

// Mean_Load_Physician_cluster not differ during $\Delta t$

19.    $i \leftarrow m$ ;

// Patient automatic assign to General physician Cluster.

20.    **Call** : Module Update
21.    **End**

## V. RESULT ANALYSIS

The simulation is done by considering the equal number of specialist physician, and the general physician number is double of any specialist physician size. The simulation is run at an estimated time for two hours i.e. 120 minutes. Two plots are presented here. Figure 1 presents the simulation with the set of physician namely Gastroenterologist, Nephrologist and General Physician. The set is composed of fifteen numbers of Cardiologists, fifteen numbers of Nephrologists and thirty numbers of General physicians. In the first set of simulation, it has been considered the patient submitted their biological report mostly related to the area of Cardio problem, Nephrology problem and other physical problem. The simulation has shown that patient assign to specialist physician uniformly up to 85 minutes i.e. one hour fifteen minutes. During this time period, some of the patient receive their medical advice and depart from the physician accepted list. It has been noticed from the simulation that after 85 minutes, there is no place to accept any new patient to the specialist physician cluster. Without any further delay, the new patient is assigned to the General physician list. Figure 2 presents the simulation with the set of the physician namely Cardiologist, Neurologist, Orthopedics and General Physician. The set is composed of fifteen numbers of Cardiologists, fifteen numbers of Neurologists, fifteen numbers of Orthopedics and thirty numbers of General physicians. The simulation presents the patient being assigned to the specialist physicians after extracting the vital features from the submitted information. The patient being assigned to the physician uniformly up to 82 minutes after that there are no ideal specialist physician or specialist physicians are heavily loaded. The entire incoming patient load after 82 minutes assigns to the General physician, who maintaining the integrity. The simulation result presented in figure 1 and figure 2 well presented the load balancing scenario. The common perception carried that number of General physician is presented as a society is large with compare the number of specialist physician. If we increase the number of General physician at random, then the patients are assigned to the specialist physician at first come first- serve basis according to the extracted feature. The remaining patient is assigned to the General physician. Obviously, that will be helpful in the case of any Natural calamities. In both the simulations, it is clearly present that any specialist physician can't be ideal if any patient requires any specialist advice according to their submitted biological data and information, the algorithm expedites the procedure.

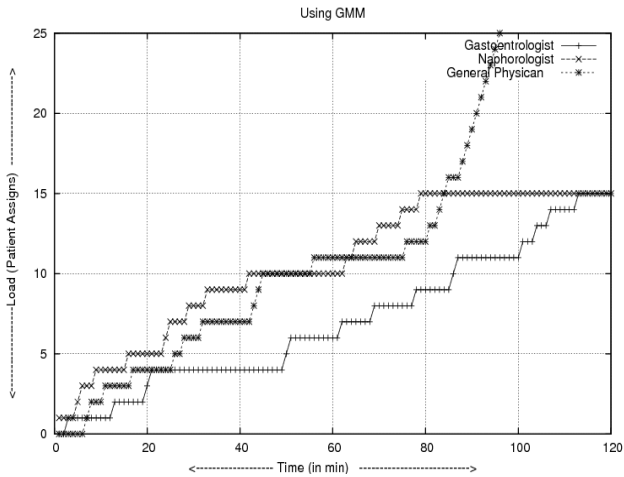

Fig. 1. Assign of patient to the three cluster of physician

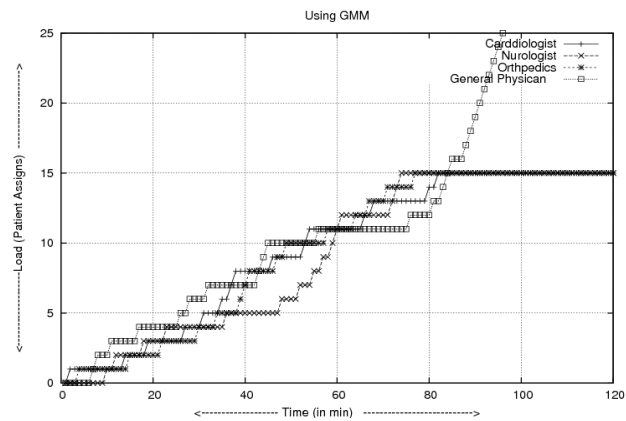

Fig. 2. Assign of patient to the four cluster of physician

VI. CONCLUSION

This work presents the fast load balancing approach for growing cluster of the patient by bioinformatics. The set of algorithms has been used to balance the patient load. If there are no spaces for the specialist physician to accept any further patient by default that patient assigns to general physician. The model has shown, the patients that needed general advice directly assign to general physician. The major modification need that at real time if any patient assigns to general physician, then the patient can't assigns to the specialist physicians in emergency, until general physicians release them. The algorithm part has to update to cope with that modification. In further the vertical handover for smart phone based patient at the client end remains further to improvements. For the massive scale of implementation in the cloud based platform being the further era in this domain.

## Acknowledgment

Author would like to thanks N K Mandal, Vidyasagar University, West Bengal India.